\documentclass[12pt]{article} 
\usepackage[hyperfootnotes=false]{hyperref}
    \usepackage{amsmath}
      \usepackage{amssymb}
 \usepackage{graphicx}
 \usepackage{xcolor}
  \newlength{\abstractwidth}
 \usepackage{mciteplus}

  \newcommand{\be}{\begin{equation}}
  \newcommand{\bea}{\begin{eqnarray}}
  \newcommand{\eea}{\end{eqnarray}}
  \newcommand{\beq}{\begin{equation}}
  \newcommand{\ee}{\end{equation}}
  \newcommand{\eeq}{\end{equation}}

  \newcommand{\half}{{1\over 2}}

\def\la{\label}

\def\32{{3 \over 2 } }

  \def\ba{\begin{eqnarray}}
  \def\ea{\end{eqnarray}}

 \def\simleq{\; \raise0.3ex\hbox{$<$\kern-0.75em
      \raise-1.1ex\hbox{$\sim$}}\; }
 \def\simgeq{\; \raise0.3ex\hbox{$>$\kern-0.75em
      \raise-1.1ex\hbox{$\sim$}}\; }

\def\nref#1{(\ref{#1})}

\begin{document}

\begin{titlepage}
  \bigskip

  \bigskip\bigskip

  \bigskip

\begin{center}
{\Large \bf {  Pure states in the SYK model \\
\bigskip
 and nearly-$AdS_2$ gravity  \\ \bigskip
  }}
 \bigskip
{\Large \bf { }} 
    \bigskip
\bigskip
\end{center}

  \begin{center}

 \bf{ Ioanna Kourkoulou$^1$ and Juan Maldacena$^2$    }
 \\
  \bigskip \rm
\bigskip
\rm
 \bigskip
 $^1$Jadwin Hall, Princeton University,  Princeton, NJ 08540, USA \\
\rm
 \bigskip
    $^2$Institute for Advanced Study,  Princeton, NJ 08540, USA  \\

  \bigskip \rm
\bigskip
 
\rm

\bigskip
\bigskip

  \end{center}

 \bigskip\bigskip
  \begin{abstract}

  We consider   pure states in the SYK model. These are given by a simple local condition on the Majorana fermions, evolved over
  an interval in  Euclidean time to project on to low energy states.  
  We find that  ``diagonal'' correlators are exactly the same as thermal correlators at leading orders in the large $N$ expansion. 
  We also describe  ``off diagonal'' correlators that decay in time, and are given simply in terms of thermal correlators. 
  We also solved  the model numerically for low values of $N$ and noticed that subsystems become typically entangled after an interaction time. 
 In addition, we identified 
    configurations in two dimensional nearly-$AdS_2$ gravity with similar symmetries. These gravity 
     configurations correspond  to states with regions behind horizons. 
     The region  behind the horizon   can be made accessible by
     modifying the Hamiltonian of the boundary theory using the the knowledge of the particular microstate. 
      The set of microstates in the SYK theory with these properties generates the full Hilbert space.

 \medskip
  \noindent
  \end{abstract}
\bigskip \bigskip \bigskip

  \end{titlepage}

   \tableofcontents


\section{Introduction}

     The SYK model involves $N$ Majorana fermions undergoing    few body interactions with random couplings \cite{Sachdev:1992fk,KitaevTalks}. 
     At low energies, it is a maximally chaotic model that has some features in common with near extremal black holes, or 
     more precisely, with nearly-$AdS_2$ gravity \cite{Sachdev:2010uj,KitaevTalks,Almheiri:2014cka,Jensen:2016pah,Maldacena:2016upp,Engelsoy:2016xyb}. 
      In this paper, we consider the evolution of particular pure states in the SYK model. 
  We   study some aspects of the thermalization of these states. We also attempt to draw some lessons for the 
  geometry associated to particular microstates for nearly-$AdS_2$ black holes. 
  
  The particular initial pure  states are obtained by combining pairs of Majorana fermions into qubit like operators and choosing 
  states with definite eigenvalues for   the $\sigma^3$ components of all qubits. By choosing different eigenvalues we get a whole basis of the 
  Hilbert space. We further evolve these states over some distance $\ell$  in Euclidean time in order to get  low energy states. 
  
Up to the first few orders in the $1/N$ expansion,  one can compute the correlators  of this model in terms 
  of finite temperature correlation functions, with $\beta = 2 \ell$. 
The correlators of fermions with the same index turn out to be the same as thermal correlators.  
 We interpret this as saying that the ``gravitational background'' for these  states is the same as that of
 the thermal state. Some other correlators, such as correlators involving fermions with different indices,  are different from the thermal ones, which are
 zero for different indices. 
 However, they can still be computed in terms of suitable  thermal correlators.  
 These are such that they decay away under Lorentzian evolution, reflecting the thermalization of the system. 
 
 The gravity dual of the SYK  model is not precisely known. 
 However, at low energies  there is an emergent reparametrization symmetry that is both spontaneously and explicitly broken \cite{KitaevTalks}. 
 This pattern of symmetries is also present in nearly-$AdS_2$ gravity, where the reparametrization symmetry is the asymptotic symmetry of 
 $AdS_2$ \cite{Almheiri:2014cka,Jensen:2016pah,Maldacena:2016upp,Engelsoy:2016xyb}.  
 In the same spirit,   we identify some nearly-$AdS_2$ gravity configurations that have properties similar to the 
 pure states in the SYK model. Namely, we will see that the symmetries are broken in a similar fashion. 
 These pure states have a gravity description which involves again the full $AdS_2$ space, but we introduce a shockwave at $t_L=0$ on the left boundary. 
 Correspondingly, in Euclidean space, we continue to have a disk, but with a special point at the boundary. 
 This special point is the source of the shock wave in the Lorentzian geometry. 
   An interesting feature of the  geometric configuration  is that it contains a region behind the horizon. The shock wave is separated from the horizon. 
 This suggests that we have a whole basis of states in the Hilbert space with  a smooth  horizon. 
 The region behind the horizon is not accesible to simple experiments by  the boundary observer. 
  However, as studied in \cite{Ahmed}, evolving the system with a modified Hamiltonian we can make some of the region behind the horizon visible. 
Here,  we can make  the whole $t=0$ spatial slice visible. 

   For this purpose he/she has to add a term to the Hamiltonian that depends on the 
 particular microstate that is chosen. The procedure is essentially the same as the one that renders wormholes traversable \cite{Gao:2016bin,Maldacena:2017axo}. 
  
  This paper is organized as follows. In section two we define the model and define a set of simple initial states. 
  In section three we display the large $N$ solution. 
  In section four we discuss some aspects of the low energy limit. 
  In section five we present some numerical diagonalization results, discussing the decay of correlators, the rise of entanglement entropy, and we check 
  statements about the large $N$ solution. 
  In section 6 we discuss some aspects of the gravity interpretation. In section 7 we give the protocol for looking behind the horizon for these states. 
   We end with a final discussion.

  \section{ Definition of the model and the initial states }

  We consider the SYK model \cite{Sachdev:1992fk,KitaevTalks}. We consider a Hilbert space generated by an even number, $N$, of  Majorana fermions 
  $\psi^i$, with $\{ \psi^i,\psi^j \} = \delta^{ij}$. 
   It has a Hamiltonian of the form 
   \be \la{Ham} 
   H = \sum_{1 \leq i<k< l< m\leq N }  j_{iklm} \psi^i \psi^k \psi^l \psi^m ~,~~~~~~~~ {\rm with} ~~~ \langle j_{iklm}^2 \rangle ={ 3!  J^2 \over N^3 }  
   \ee
    with couplings
  $j_{iklm}$ which are all independent random numbers drawn from a gaussian distribution with variance set by $J$, see \cite{Maldacena:2016hyu} for more details.   
More generally one can also consider a model with a Hamiltonian involving $q$ fermions $H 
=i^{q\over 2}   \sum_{k_1\cdots k_q}  j_{k_1 \cdots  k_q} \psi^{k_1} \cdots \psi^{k_q} $. 
  
      We are interested in considering the evolution of   special pure states. For example, we can consider the state 
  $|B\rangle$ that obeys the conditions 
  \be \la{DefB} 
   (\psi^1 - i \psi^2 )  |B\rangle =0 ~,~~~~~~(\psi^{2k-1} - i \psi^{ 2 k } )  |B\rangle =0 ~,~~k=1,\cdots , N/2 
   \ee
   If we imagine $\psi^1$ and $\psi^2$ as proportional to the $\sigma^1$ and $\sigma^2$ Pauli matrices, then $|B\rangle$ will have spin minus under 
   $\sigma^3$.  More precisely, we can say that the state $|B\rangle $ has all plus  eigenvalues for the operators 
   \be
   \la{Sk} 
   S_k \equiv  2  i \psi^{2k-1} \psi^{ 2 k}  ~,~~~~~~~~S_k^2 =1 ~,~~~~~k=1,\cdots , N/2
   \ee 
   
   More generally, we can define a whole set of states $|B_s\rangle $ by the conditions 
   \be \la{Bs}
   (\psi^{2k-1} - i s_k  \psi^{ 2 k } )  |B_s\rangle=0  ~,~~~{\rm or} ~~ S_k |B_s\rangle = s_k |B_s \rangle ~,~~~~{\rm with}~~s_k = \pm 1
   \ee 
    This defines $2^{N/2}$ states, one for each choice of the 
   signs of all the $s_k$,  which form a basis of the Hilbert space.

   The SYK evolution with \nref{Ham} will   give states which are linear combinations of these states. 
   At long times we expect to get fairly generic linear combinations so that the state becomes effectively thermalized (even though it remains a pure state).
     If we think of $|B\rangle$ as a simple state were all qubits point up, then the 
   evolution will start flipping some of the qubits so that we start getting a more general superposition of states with qubits pointing up and down\footnote{
   The SYK evolution does preserve the sign of $\prod_{k=1}^{N/2} S_k = (-1)^F$.}. More importantly,  
    we get linear superpositions of such states. The SYK evolution can be viewed
   as a set of simple quantum gates that acts on the simple state generating a more complex state.

  We expect these states, \nref{Bs}, to have an  energy close to zero for the SYK hamiltonian. These are states of high energy compared to the minimal energy 
  of the SYK model which is of order $E_0 = - ({\rm number} ) N $. 
  
  We can produce lower energy states by evolving with the Euclidean Hamiltonian $|B(\ell) \rangle =e^{ - \ell H } |B \rangle$. In this way we can form an overcomplete set 
  of low energy states.  
 We expect that the expansion of $|B(\ell) \rangle $ is in terms of the energy eigenstates is 
  \be
   |B(\ell) \rangle \sim \sum_\alpha e^{ - \ell E_\alpha } c_\alpha |E_\alpha \rangle  \la{EnEx} 
   \ee
   where the typical $|c_{\alpha }|^2 $ is of order of $2^{ - {N \over 2} +1 }~$   \footnote{ Of course, some of the $c_\alpha$ are zero 
   for symmetry reasons. For example, since $(-1)^F$ commutes with the Hamiltonian, all states that appear have to 
   have the same value of $(-1)^F$ as the state $|B\rangle$. }. In figure \ref{RandomCs}
    we see an example for $N =30$ obtained by exact numerical diagonalization. 
  
 Note that we can  average the correlators over all choices of signs $s_k$. This reproduces the thermal ensemble exactly, 
             \be \la{ThEn}
             \sum_{s_k} \langle B_s(\ell)  | \psi \cdots \psi |B_s(\ell) \rangle = Tr[ e^{ - \beta  H } \psi \cdots \psi ] ~,~~~~~ \beta = 2 \ell 
             \ee
              This means that we can view 
             the  states   $|B_{s_k}(\ell) \rangle$  as an  (overcomplete) basis of the low energy states relevant to the dynamics at temperature $\beta$.  
             We see that after averaging over all sign choices $\{ s_k\}$ for the states $|B_s(\ell) \rangle$  we get the exact thermal average. 
   Of course, \nref{ThEn} is not at all surprising, given the way we have defined the states. What is more interesting is that each
     { \it individual } state $|B_s(\ell) \rangle $ gives 
   rise to correlators that look thermal to high accuracy, as we will demonstrate below.           
  
             These ``boundary states'' are the one dimensional analog of similar boundary states that were used in 
             \cite{Calabrese:2005in,Cardy:2017ufe} to model quenches
             from vacua of gapped short range  Hamiltonians to CFTs in 1+1 dimensions.  Related quantum quenches   in the SYK model 
             are discussed in \cite{Eberlein:2017wah}.

  \section{ The large $N$ solution } 
  
  \subsection{Two point functions from thermal ones} 
  
  In this section we analyze this problem in the large $N$ limit. The proper way to treat the average over 
  random couplings is to introduce replicas \cite{Maldacena:2016hyu}. 
  However, one finds that (for the replica diagonal solution) the interaction between replicas is down by $1/N^3$ (or $1/N^{q-1}$ more generally). 
  This means that, to leading orders in $N$, we can treat the couplings $j_{iklm}$ as time independent gaussian fields with the two point function in 
  \nref{Ham}. In this approximation, the model has an $O(N)$ symmetry. A particularly interesting subgroup of $O(N)$ is the one that flips the sign of any
  of the even indexed fermions. For example, we can consider the element that flips the sign of $\psi^2$ leaving the rest unchanged. There is another element
  that flips the sign of $\psi^4$, etc. We call this collection of $O(N)$ group elements, the  ``Flip Group''  (it is  generated by 
  reflections along the even directions). 
  The boundary states $|B_s\rangle$ \nref{Bs} are {\it not} invariant under these elements. The element 
  that flips the sign of $\psi^2$, changes the sign of $s_1$ in $|B_s\rangle$, so it maps one possible state into another.  We can think of this as flipping the 
  sign of the first spin. 
   Notice that  two point functions such as $\psi^1(\tau_1) \psi^1(\tau_2)$ or $\psi^2(\tau_1) \psi^2(\tau_2)$ are individually
    invariant under the ``Flip Group''. We call such two point function ``diagonal'' two point functions.  
 Diagonal  correlators have the same values in all $|B_s\rangle $ states. 
 On the other hand we have also shown that the average over all states $|B_s\rangle$ is the same as the thermal average \nref{ThEn}. 
 This means that these two point functions have a value with is identical to their thermal averages. 
 The same argument also implies that the following overlap is given in terms of the partition function 
 \be
  \langle B_s | e^{ - 2 \ell  H } |B_s \rangle = 2^{ - N/2} Z(\beta) ~,~~~~~~~~\beta = 2 \ell  \la{Ovlp}
 \ee
 since the same argument indicates that it is independent of $s_k$, to leading order in the $1/N^{q-1}$ expansion. Note that both sides of \nref{Ovlp} also 
 have a $1/N$ expansion. 
 This  equation says that \nref{ThEn} holds to good approximation  for each state, not just in average. This is 
 the same self averaging that we can invoque regarding the random couplings, but now applied to the initial state\footnote{ These two self averages are related if 
  we view the initial state as arising from a long Euclidean time evolution with random cuadratic couplings.}.

 The operator $ \psi^1(\tau_1) \psi^2(\tau_2)$ is not invariant under a reflection of the $\hat 2$ axis.
 However,  we can consider the 
 combination $   \psi_1 (\tau_1) \psi_2(\tau_2) S_1 $ which is  indeed invariant. 
 Note that 
 \bea
 & ~& \sum_s \langle B_s| \psi^1(\tau_1) \psi^2(\tau_2)  s_1 |B_s\rangle =  \sum_s \langle B_s| \psi^1(\tau_1) \psi^2(\tau_2)  2  i \psi^1(0) \psi^2(0) |B_s\rangle=
  \cr
 & ~& ~~=
 2i  Tr[ e^{ - 2 \ell H }  \psi^1(\tau_1) \psi^2(\tau_2)    \psi^1(0) \psi^2(0) ] \sim - 2 i G_\beta(\tau_1) G_\beta(\tau_2)  \la{TwoAvg}
 \eea
Since $   \psi_1 (\tau_1) \psi_2(\tau_2) S_1 $ is invariant under the ``Flip Group'',   \nref{TwoAvg}  also shows that the final result also holds over element 
 of the first sum, for each state $|B_s\rangle$. 
 
 In conclusion,  defining the  two point functions in the state $|B_s\rangle $ as 
 \bea
  G_{\rm diag} (\tau, \tau')  &=& {  \langle B_s(\ell) | \psi^i(\tau_1) \psi^i(\tau_2) |B_s(\ell) \rangle \over  \langle B_s(\ell) |B_s(\ell) \rangle} =
 ~,~~~~~~~~~~~~~{\rm no~ sum}  \la{RatioNat} \\
  &=&  { \langle B_s |e^{ - ( 2 \ell - \tau_2 ) H } \psi^i e^{ - (\tau_2 - \tau_1) H } \psi^i  e^{ - \tau_1 H } 
 |B_s \rangle 
 \over  \langle B_s | e^{ - 2 \ell H }  |B_s \rangle } ~,~~~\tau_2 > \tau_1~,~~~~~~~~{\rm no~ sum} 
 \cr \la{OffDiag}
  G_{\rm off}(\tau,\tau') &= & s_k { \langle B_s(\ell)  | \psi^{ 2k-1} (\tau_1) 
 \psi^{2 k } (\tau_2) |B_s(\ell) \rangle \over  \langle B_s(\ell) |B_s(\ell) \rangle}  ~,~~~~~~~{\rm no~ sum} 
 \eea
 we find that 
 \bea
   &~&   G_{\rm diag} (\tau,\tau')= G_{\beta}(\tau-\tau') ~,~~~~~~~~~~
  \beta = 2 \ell   ~~~
       \label{Twoptdi} 
\\
&~&     G_{\rm off}(\tau,\tau' ) = 2i \langle \psi^1(\tau )\psi^2(\tau') \psi^1(0 )\psi^2 (0)\rangle_\beta = -2 i G_{\beta}(\tau) G_{\beta}( \tau' ) + 
o ( 1/N)   ~,~~~~~~~  \label{Twoptoff}
       \eea
  These results are valid at leading   order in the  $1/N^{q-1}$ expansion, but the last equality in \nref{Twoptoff} is valid only to leading order in the $1/N$ expansion. 
  In principle, it is 
   possible to add the first $1/N$ correction to \nref{Twoptoff} that comes from the connected part of the  four point function. 
   Notice that part of the statement is that in \nref{RatioNat} \nref{OffDiag} there
  is no dependence on $i$ or $k$ or the set of $s_k$. In the second line of \nref{RatioNat} 
  we have assumed that $\tau_1 < \tau_2$   (otherwise it needs to be reordered).


  \subsection{Comments}
  
     \begin{figure}[h]
\begin{center}
\includegraphics[scale=.45]{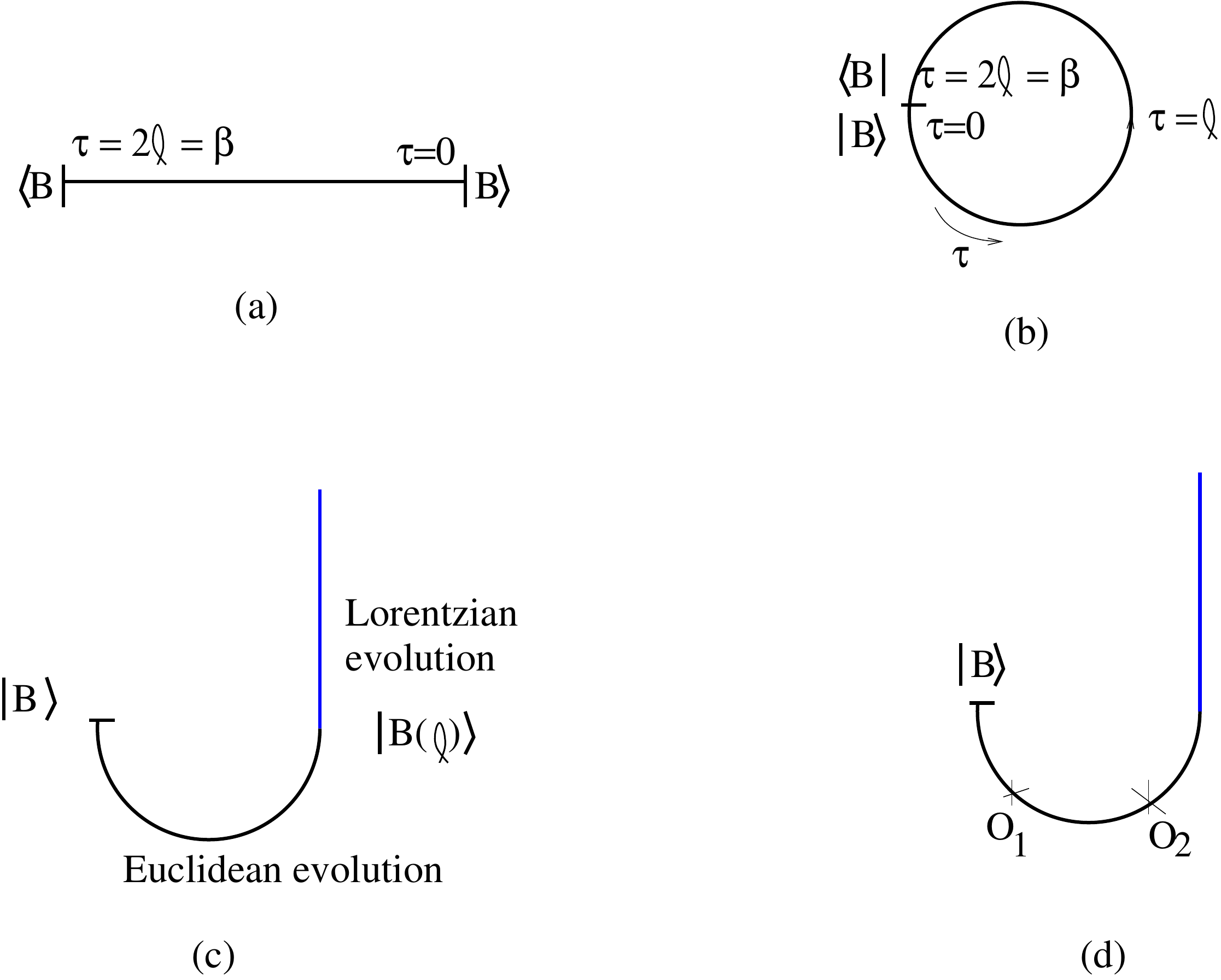}
\caption{(a) The Euclidean computation pictured as an interval of size $\beta = 2 \ell$. (b) The same Euclidean computation pictured as a circle, with a special point 
where we project on to the state $|B\rangle$.   (c)  Evolving by  Euclidean time $\ell$ we get the state $|B(\ell)\rangle$, which we 
can then evolve using  Lorentzian evolution. (d) By inserting simple 
operators $O_1$, $O_2$,  at intermediate Euclidean times we can get states containing some small excitations.   }
\label{Circle}
\end{center}
\end{figure}
    
      Let us note the following points 
  
  \begin{itemize} 
  
  \item  The diagonal correlators are {\it exactly} thermal at large $N$.
   In particular, they only depend on the difference of times, despite the presence of the 
  boundary at $\tau=0, 2 \ell$. If we know the numerical or analytic (e.g. at large $q$) large $N$ solution of the finite temperature model,
   then \nref{Twoptdi}, \nref{Twoptoff} give us a direct solution for
  the pure state problem. 
  
  \item Note that there is no singularity in $G_{\rm off} $ at $\tau = \tau'$ since the two different fields anticommute. 
  
   \item \nref{Twoptdi} \nref{Twoptoff} obey the boundary condition $G_{\rm diag}(\tau,0) = - i G_{\rm off}(\tau,0)$ implied by \nref{DefB}, after using the 
  UV form of the thermal correlator, $G_{\beta}(0)=\half$. 
  
  \item  Douglas Stanford has checked   \nref{Twoptdi} \nref{Twoptoff}     against a direct numerical solution of the large $N$ 
  Schwinger Dyson equations with the appropriate boundary conditions \cite{DouglasPrivate}. 
  
  \item The diagonal correlators are independent of the state we started from. They are independent of the choice of $s_k$ in \nref{Bs}.

  \item The off diagonal correlators are non-zero and they depend on the precise initial state we start from, through the signs $s_k$. 
  
  \item Notice that these formulas are valid for all values of $\beta J $, or $\ell J$,    to leading order in the $1/N^{q-1}$ expansion (for $ \beta J \ll N $).  
  
  \item For $q\geq 4$,    \nref{Twoptdi} at order $1/N$ implies that all the operators that appear in the OPE of $\psi^i \psi^i$ also have the 
   expectation values  as in the thermal state. 
  
  \item If we interpret the diagonal correlators as giving rise to   the full  ``gravity''  background, then this background is exactly the same as the one we have
  for the thermal state, or the thermofield double. 

    \item We can set $\tau = \ell + i t$ to get the Lorentzian correlators at time $t$  in the state $|B(\ell) \rangle $.  In particular, note that the thermal correlator 
    $G_\beta( { \beta \over 2} + i t) $ is real, so that \nref{Twoptoff} is consistent with the anti-hemiticity of the operator \nref{OffDiag} when $t=t'$. 
   
   \item We can view  the Euclidean interval as   a full circle with a point where the states running along the circle are projected to
   joint eigenstates of the operators $S_k$, see figure \ref{Circle}(b).

    \item By inserting operators at  Euclidean times $ 0 < \tau_i < \ell$, we can get other ``close-by'' states, see figure \ref{Circle}(d). 

\item The classical action on the solution gives us the overlap \nref{Ovlp}, and it coincides with the partition function 
up to a constant, as indicated in \nref{Ovlp}.

    \item The following is a side comment.
     One could imagine starting with the  SYK model  in \nref{Ham} and then attempt to introduce a time dependence by 
    taking a Hamiltonian which is $H_{\rm new} = g(t) H$, where $H$ is in \nref{Ham}. However, this time dependence can be completely removed by redefining 
    the time to $\tilde t$, via   $d\tilde t = g(t) dt $. In terms of the time $\tilde t$ we have the standard time independent Hamiltonian.  Of course, if we were to change
    individual couplings relative to each other, that can change the physics.   
   
    \item Suppose we define the ratio 
    \be \la{RatioAvg} 
    {  \langle B_s(\ell) | \psi^i(\tau_1) \psi^i(\tau_2) |B_s(\ell) \rangle \over  Tr[e^{ - \beta H} ] }
    \ee
     instead of the more natural one
    in \nref{RatioNat}. We can  now argue that the average over couplings of this new ratio gives exactly the same as the average over couplings of the thermal correlator. 
    The reason is that when we compute this ratio using the replica trick, we impose the $|B\rangle$ boundary condition on one replica but the thermal one on the rest. 
    Then the same symmetry argument we used to get \nref{Ovlp} is valid for the replicated problem. We checked this for $N=24$ in figure \nref{CouplingsAvg}.

   \end{itemize}

  \section{ Low energy limit , almost conformal limit} 
  
 In this section we will discuss some  low energy aspects of the above formulas \nref{Twoptdi} \nref{Twoptoff}. We consider $1 \ll \ell J \ll N $ so that we 
 go to the almost conformal limit. In this case we can use the conformal limit of the thermal correlators \cite{Sachdev:1992fk,KitaevTalks,Maldacena:2016hyu}
 \be \la{theE}
     G_{\beta}(\tau ) = { c_\Delta  \over \left[ { J \beta \over \pi }  \sin { \pi \tau \over \beta } \right]^{ 2 \Delta } }  ~,~~~~~~~
     c_\Delta \equiv  \left[ \left( \half - \Delta \right) { \tan \pi \Delta \over \pi } \right]^\Delta
    \ee
    where $\Delta =1/4 $ for the Hamiltonian in  \nref{Ham}\footnote{When $H \sim \psi^q$ we get $\Delta =1/q$  \cite{Maldacena:2016hyu}.}.  
  We get  the off diagonal correlator in Lorentzian time  by setting $\tau = { \beta \over 2} + i t $ in \nref{theE}
  \be \la{GoffLow}
  G_{\rm off}(t,t')  = - 2  i  { c_\Delta^2 \over \left[\left( { J \beta \over \pi } \right)^2 \cosh { \pi t \over \beta } \cosh { \pi t' \over \beta } \right]^{ 2 \Delta } }   ~,~~~~~~ \beta = 2 \ell 
  \ee
  The correlator looks like  the product of two thermal thermal correlators and it
   decays in time as expected for a system that is thermalizing. Notice that the decay time is of the order of
   the temperature. 
  
  It is instructive to consider the small euclidean time limit of the correlators (or large $\ell$) to obtain 
  \be \la{Powers}
  G_{\rm diag}=   { c_\Delta \over | J (\tau - \tau')|^{ 2 \Delta } } ~,~~~~~~~~~~G_{\rm off}(\tau,\tau')= - 2 i { c_\Delta^2  \over|J^2 \tau \tau' |^{ 2 \Delta } }
  \ee
  These are the correlators in Euclidean time, close to the boundary state $|B\rangle$. 
  We see that we {\it cannot} check the boundary condition \nref{DefB} purely within the low energy limit \nref{Powers}, since the $\tau \to 0$ limit of the 
  conformal limit of $G_{\rm off}$ gives infinity. This is not a problem, it is merely saying that in order to check the boundary condition we need to evaluate the 
  first factor of $G_\beta(\tau)$ in \nref{Twoptoff} at zero, and we should use the short distance form of the exact solution which is $G_{\beta}(0) =1/2 $. 
  This then allows us to check the boundary condition, which is of course obeyed already at the level of \nref{Twoptoff}.

  In the thermal case, we develop an emergent reparametrization symmetry that is also explicitly broken by the 
  Schwarzian action \cite{KitaevTalks,Maldacena:2016hyu}. We expect something similar  in our problem. 
  One difference is that there is a special point at $\tau =0$ (and $\tau= 2\ell$) where the boundary sits. So we expect that the reparametrization 
  mode, $\tau \to \varphi(\tau)$, 
   should obey the boundary condition $\varphi(0)=0$, $\varphi(2\ell) = 2 \ell$. 
   In addition, it also needs to obey $\varphi'(0)=1$, $\varphi'( 2 \ell) =1$. Namely, we should fix its gradient at the position of the 
  boundary. Indeed, if we define 
  \be 
  \la{ReparOff} 
  G_{\rm off , ~ \varphi}(\tau,\tau') = [ \varphi'(\tau) \varphi'(\tau') ]^\Delta G_{\rm off }( \varphi(\tau) - \varphi(\tau') ) 
  \ee
   then the boundary condition at zero will be 
  obeyed only if $\varphi'(0) =1$. 
  
  We know that the thermal solution spontaneously breaks the reparametrization symmetry to $SL(2)$. The boundary conditions imply that only one 
  element of $SL(2)$ survives. Indeed, out of the three conformal  Killing vectors on the circle: $1 , \cos \tau , \sin \tau $ (for $\beta = 2 \pi $), only the combination 
  $\zeta = 1 - \cos\tau $ remains as a symmetry.  Importantly, this is a symmetry also of the off diagonal correlator $G_{\rm off}$. The other two $SL(2)$ generators are not a symmetry of $G_{\rm off}$. 
  
  We mentioned above that it is useful to think of the interval $\tau \in [  0 , 2 \ell ] $ as a circle with a special point, see figure \ref{Circle}(b).
   It is useful to send this special point to 
  infinity and map the circle to a line. Explicitly, we map  $t_{\rm line} =-{ \pi } \left[  \beta  J^2  \tan { \pi \tau \over \beta }  \right]^{-1} $. 
  Under this map, the off diagonal correlator becomes simply a constant, 
  \be
  G_{\rm off }(t,t') = -2 i { c_\Delta^2 }  \la{GoffLine}
  \ee
   And the surviving element of $SL(2)$  is simply 
   translations along this line, $t_{\rm line} \to t_{\rm line } + $(constant). 
  
   The Schwarzian action on the circle becomes 
   \be \la{Sch}
   S = - { \alpha_S N \over J } \int d \tau  \{   \tan { \varphi(\tau)   \pi \over \beta }  , \tau  \}  ~,~~~~~{\rm with}~~~\{ f(\tau) ,\tau\} \equiv \left( { f'' \over f' } 
   \right)' - \half \left( { f'' \over f' } \right)^2 
   \ee
   where $\alpha_S$ is a numerical constant \cite{Maldacena:2016hyu}, with boundary conditions
    $\varphi(0)=\varphi(2\ell) =0$, $\varphi'(0) = \varphi'(2\ell) =1$. 
 
In accordance with \nref{Ovlp}, the one loop correction is the same as for the thermal partition function. 
The boundary conditions for the Schwarzian variable are removing two of the  $SL(2)$ zero modes. 
 They are still leaving one zero mode.  But we should   not integrate over  any of these zero modes  anyway,  so we get the same answer.

      \section{ Some exact diagonalization results } 
  
  In this section we present some results obtained by exact diagonalization of the   Hamiltonian \nref{Ham} for some
  values of $N$. 
  
  \subsection{The pure states as a typical state in Hilbert space} 
  
\begin{figure}[h!]
\begin{center}
  \includegraphics[scale=.37]{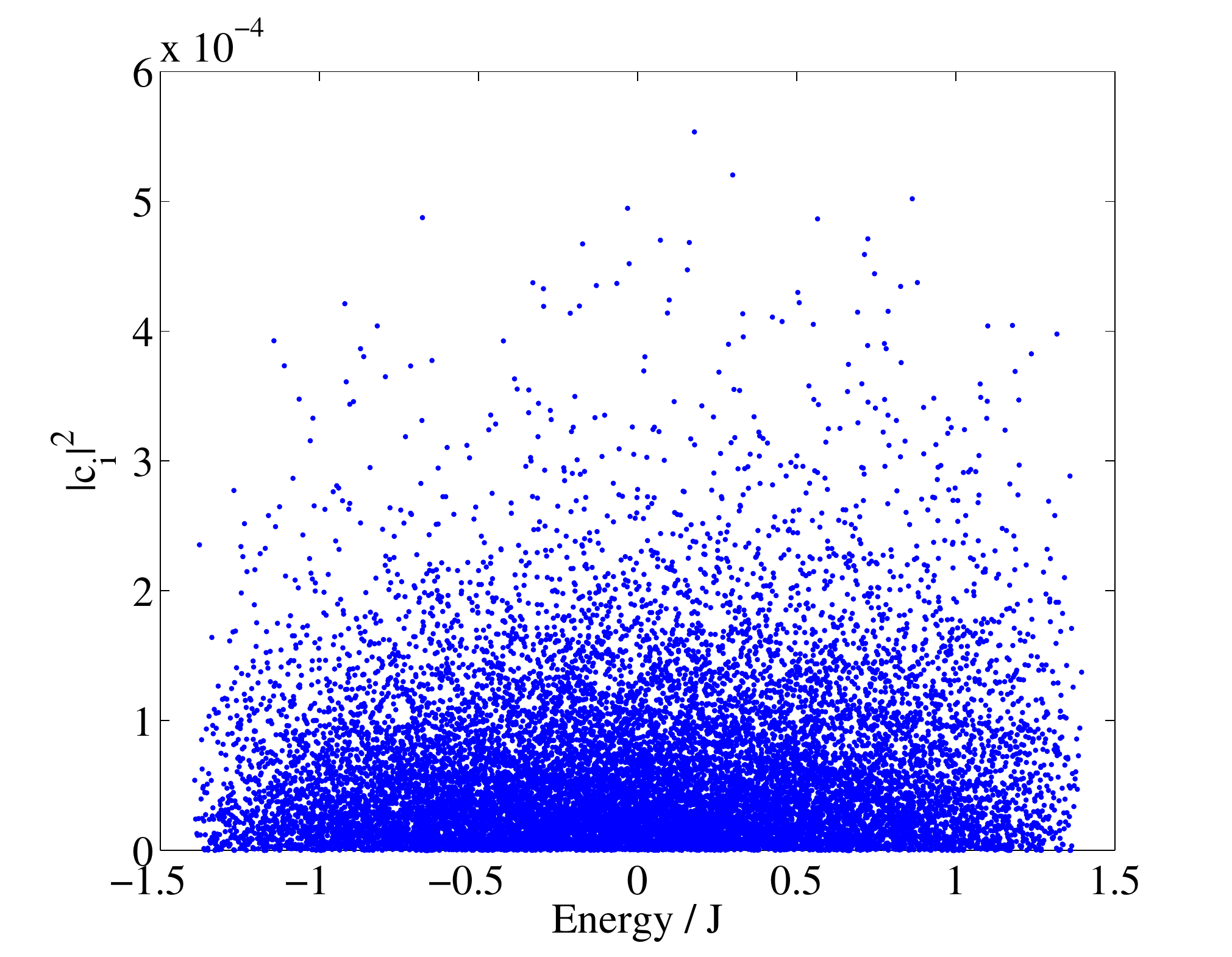} ~~ \includegraphics[scale=.37]{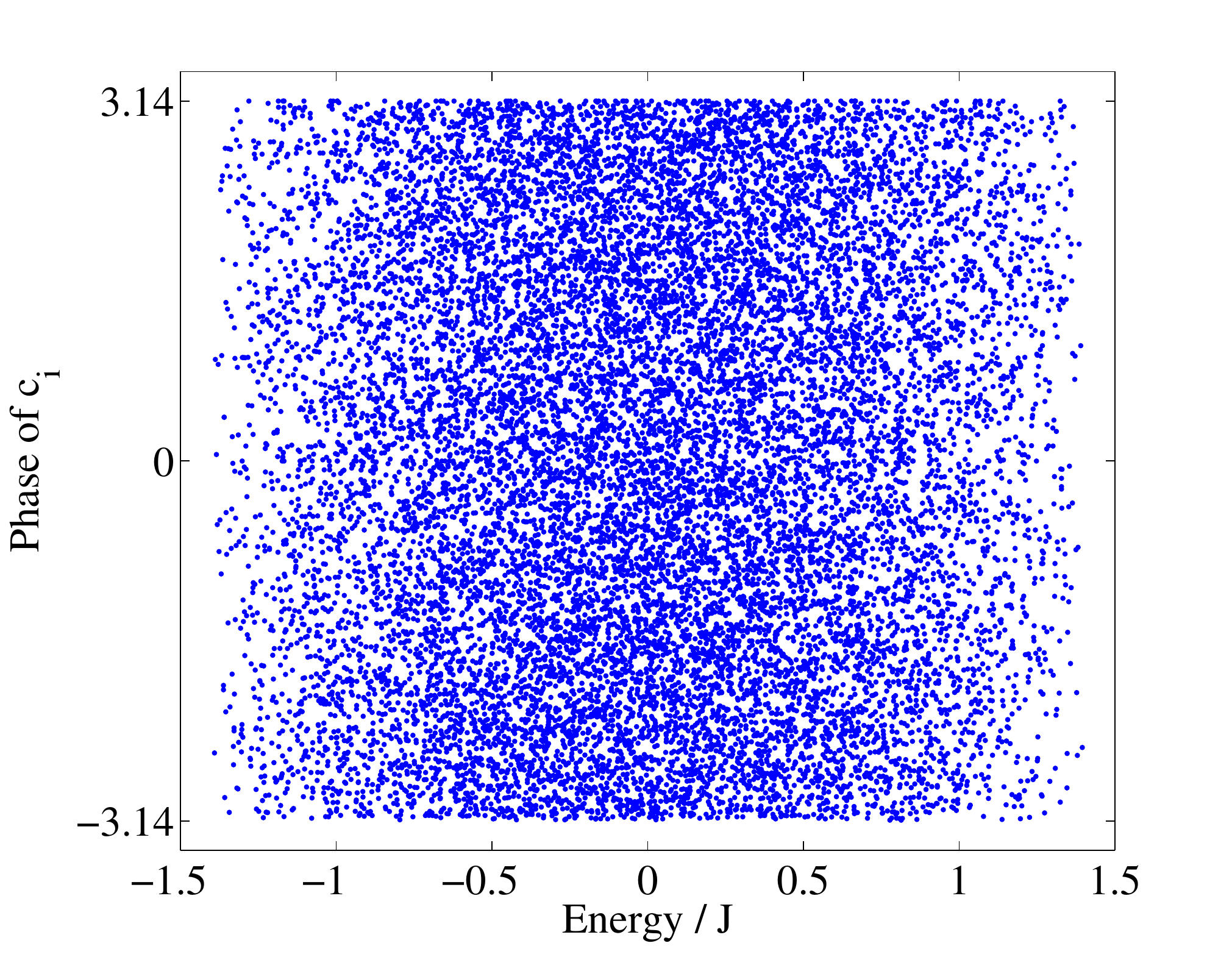}
  \caption{ We set $N=30$. In the left we    plot the square of the coefficients of the non-zero expansion of $|B\rangle$ in terms of the energy of the energy eigenstates. 
(Half of the coefficients are automatically  are zero due to the $(-1)^F$ symmetry). They are random looking.
 The average value of the square of the non-zero coefficients 
is $2^{-N/2 +1}$, which is about $ 0.6  \times  10^{-4}$. Note that the density of eigenstates is not uniform along the horizontal axis. 
On the right see the   phases of the coefficients.   More precisely, since the phases of the energy eigenstates can be chosen independently for each eigenstate,
 we really plot the difference in phases for
  two different states $|B_s\rangle $, $|B_{s'}\rangle$.  }
\label{RandomCs}
\end{center}
\end{figure}

  First we consider the state 
   $|B\rangle$ \nref{DefB} for $N=30$  with no further Euclidean 
  evolution.  We expand the state in terms of energy eigenstates as in  \nref{EnEx}. Up to constraints given by the discrete symmetries, we 
  get random looking values for $c_\alpha$, see figure \nref{RandomCs}.  The average energy for this state is close to zero, as expected. 
  This is a relatively high energy state compared to the minimum energy state. 
  
  \subsection{ Correlation functions } 
  
\begin{figure}[h]
\begin{center}
 \includegraphics[scale=.5]{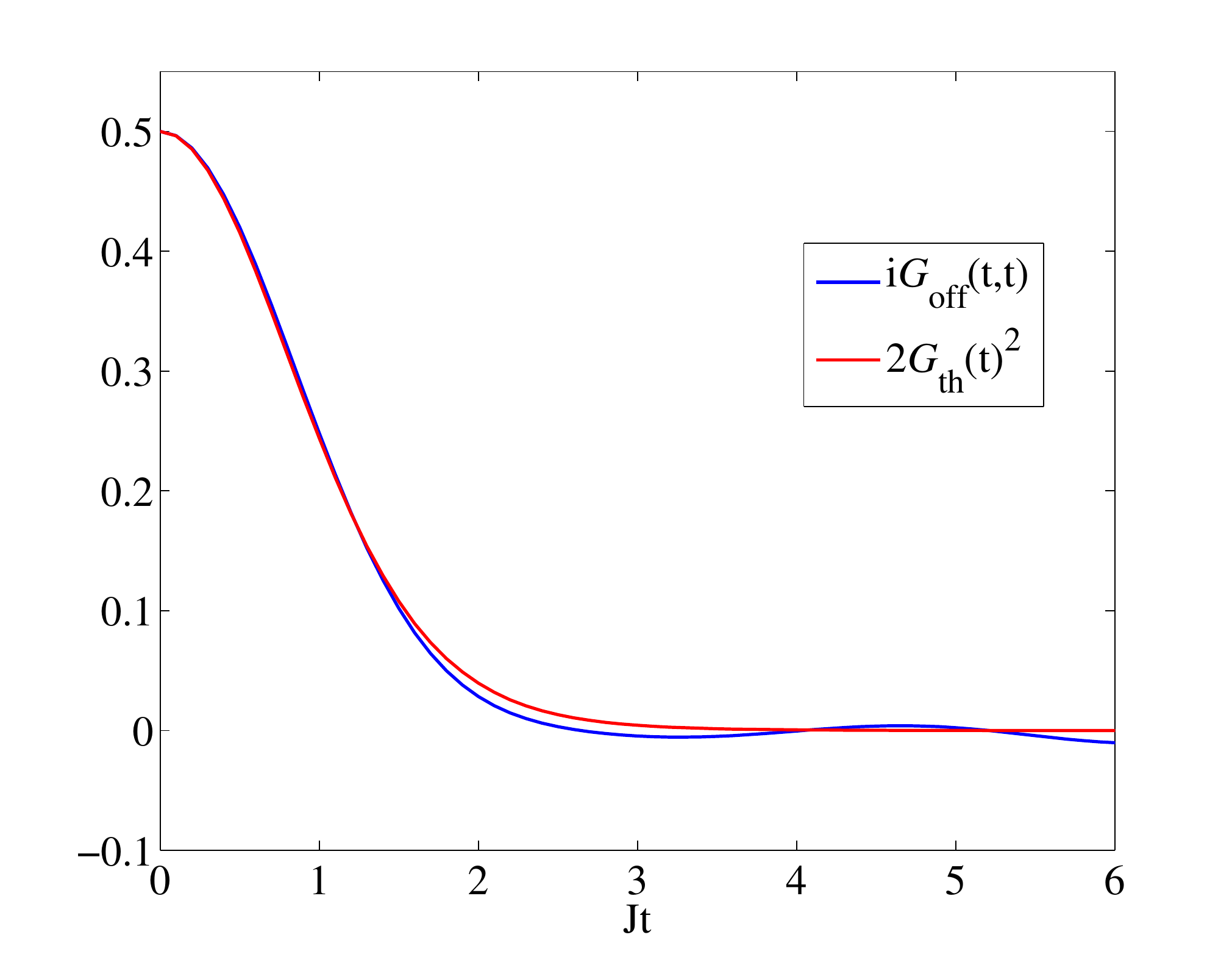}
\caption{ We set $N=24$.  We plot   the Lorenzian time expectation value of the operator $S_1(t)/2 = i G_{\rm off}(t,t)$
 on the state $|B\rangle$  \nref{Bs}  (with $\ell=0$). 
We see that it decays over at time of the order of the interaction time $1/J$. We also plot twice the square of the thermal correlator 
at $\beta =0$. We see that \nref{Twoptoff} holds pretty closely despite the low value of $N$.    }
\label{Goff}
\end{center}
\end{figure}

  We consider the expectation value of the operator $S_1(t)/2  =   i \psi^1(t) \psi^2(t)$ evaluated in Lorentzian time 
  on the state $|B\rangle$. See figure \ref{Goff}. 
   We see that it decays to zero as we expect for a thermalizing system. But it has small 
  oscillations as we expect in a unitary theory. In a unitary theory this correlator cannot decay to zero for all 
  initial states because a suitable linear combination of initial states should be able to give us a state with 
  eigenvalue $S_1 =1$ at any time. For a detailed analysis of the long time behavior in SYK see 
  \cite{Cotler:2016fpe}. We have also compared the answer to the twice the  square of the thermal correlator, as predicted by 
  \nref{Twoptoff}. They match fairly closely despite the relatively low value of $N$. 
    
  \subsection{Entanglement entropy of subsystems }

     \begin{figure}[h]
\begin{center}
 \includegraphics[scale=.6]{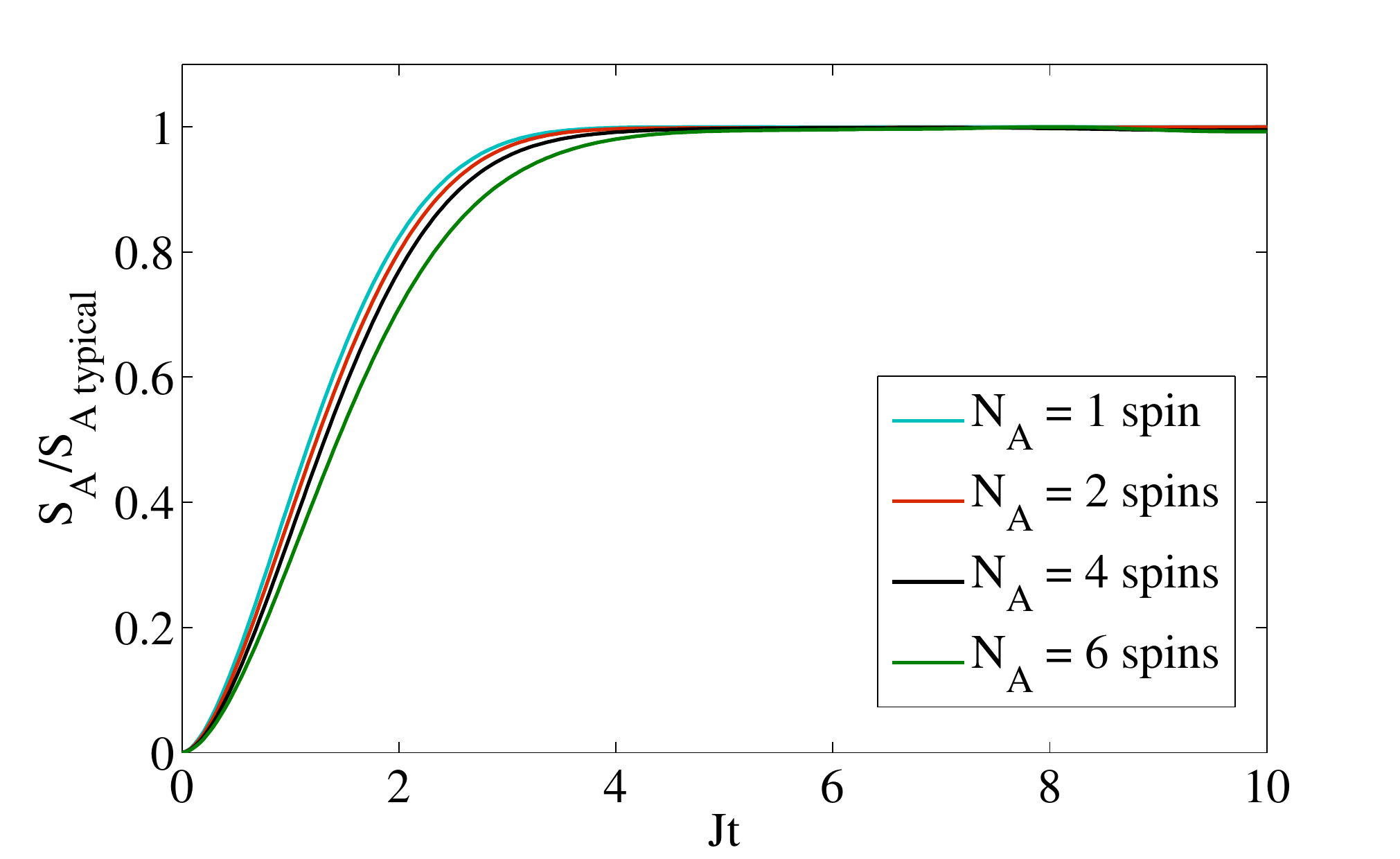}
\caption{ Here $N=24$. 
Ratio of the entanglement entropy of a subfactor of $N_A$ spins (or $2 N_A$ Majorana fermions), to the entanglement entropy for a typical 
random state in the Hilbert space  \cite{Page:1993df}. Something to note is that the time it takes to saturate is 
independent of the size of the subsystem. This is different from a local spin chain and it reflects the all to all nature of
the SYK Hamiltonian.   }
\label{Entanglement}
\end{center}
\end{figure}

  Here we consider the boundary state $|B\rangle$ and we  evolve it in Lorentzian time. We can organize the Hilbert space as a tensor product 
  of qubits, viewing the first qubit as the one whose $\sigma^3$ is given by $ S_1 =  2 i \psi^1 \psi^2$, and similarly 
  with the other qubits. More precisely, we represent the $\psi^i$ in terms of qubits by a Jordan Wigner transformation\footnote{Explicitly: 
$ \psi^{2k-1} = \sigma^1_k \prod_{i=1}^{  k-1 } \sigma^3_i ~,~~~\psi^{2k} = \sigma^2_k \prod_{i=1}^{  k-1 } \sigma_k^i $.} , and then we look at the 
  tensor decomposition of the Hilbert space in terms of the Hilbert spaces of each of these qubits. 
   The initial state is simple in terms of these qubits because it obeys a condition
   $S_k |B\rangle = |B\rangle $. It is a factorized state. 
   However, the evolution by the SYK Hamiltonian gives us general linear combinations of states with 
   definite spins.  So if we consider the 
   subfactor of the Hilbert space generated by the first few qubits, the initial state is unentangled with the rest, but
   it will become entangled under time evolution. In fact, it becomes rapidly entangled as in a typical state 
   of the Hilbert space \cite{Page:1993df}.
     By rapidly, we mean a  time of order $1/J$, which is the characteristic interaction time. Interestingly, the
   time  to reach the typical entanglement is independent of the size of the 
   subsystem\footnote{We thank D. Stanford for emphasizing this point to us.}. This fact was demonstrated analytically at large $N$ in 
   \cite{Magan:2017udh}, by  showing that the density matrix for a subset of spins factorizes at large $N$ (see also \cite{Magan:2015yoa,Magan:2016ehs,Magan:2016ojb}). 
    In figure \nref{Entanglement}  we see a plot of the ratio of these entanglement 
   entropies to the values we expect for a typical state in the Hilbert space, as computed in \cite{Page:1993df}. 
    For $N\gg 1$, this typical entanglement entropy is close to maximal, $S_A \sim \log N_A$, where $N_A$ is the number qubits of the subsystem. 
   However, at finite $N$ the typical entanglement  is slightly less than maximal. It is given by a formula  $S_{\rm typical}(N_A,N-N_A)$,
    computed in \cite{Page:1993df}. We see in figure \nref{Entanglement} that the evolution of entanglement is fairly independent of the subsystem size and that
    it reaches the maximal value at the same time for all subsystem sizes. 
   Similar subsytem entanglement entropies were computed for the ground state in \cite{Fu:2016yrv}.

  \subsection{Euclidean correlators at finite $N$ }

     \begin{figure}[h]
\begin{center}
 \includegraphics[scale=.35]{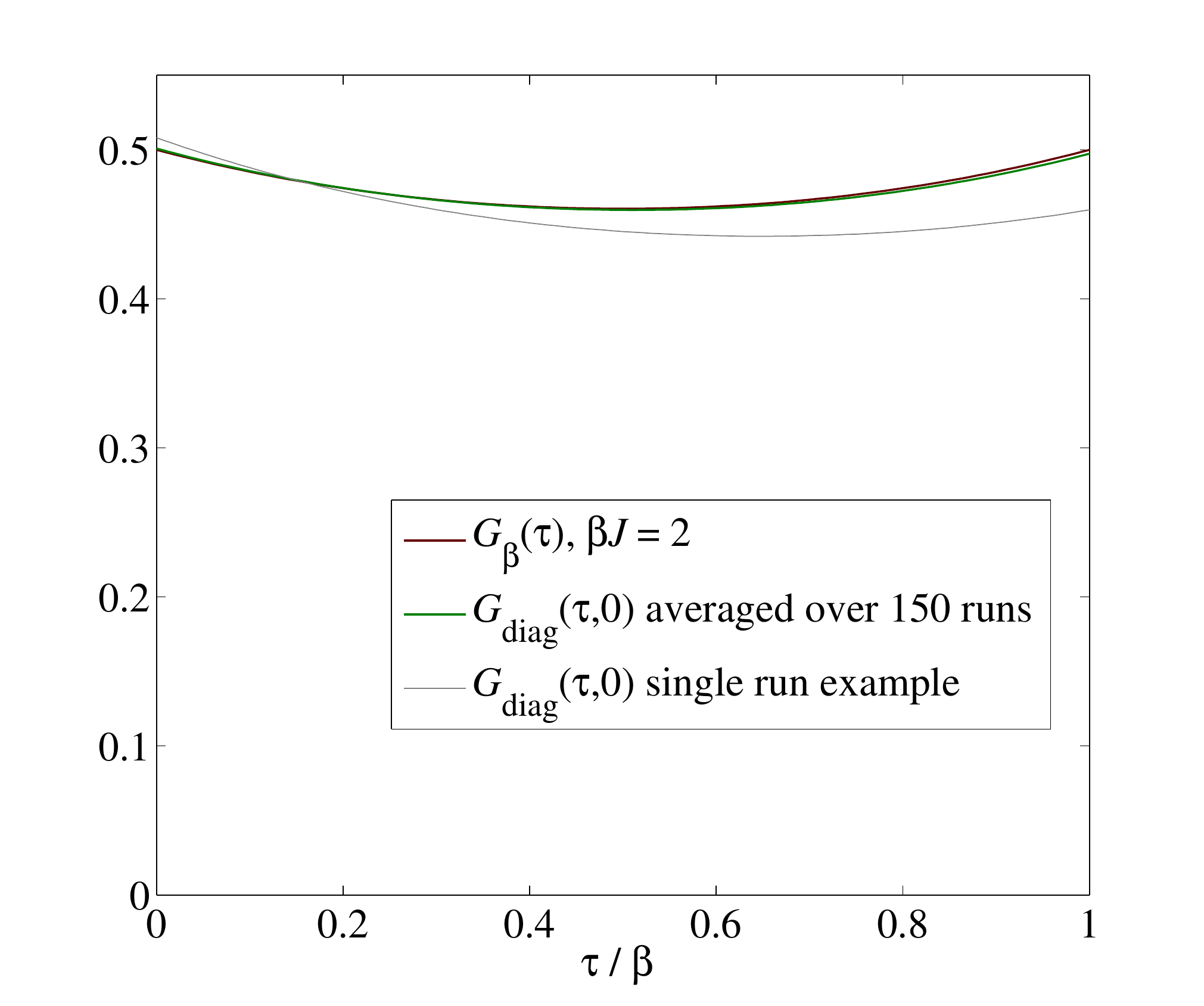}  \includegraphics[scale=.4]{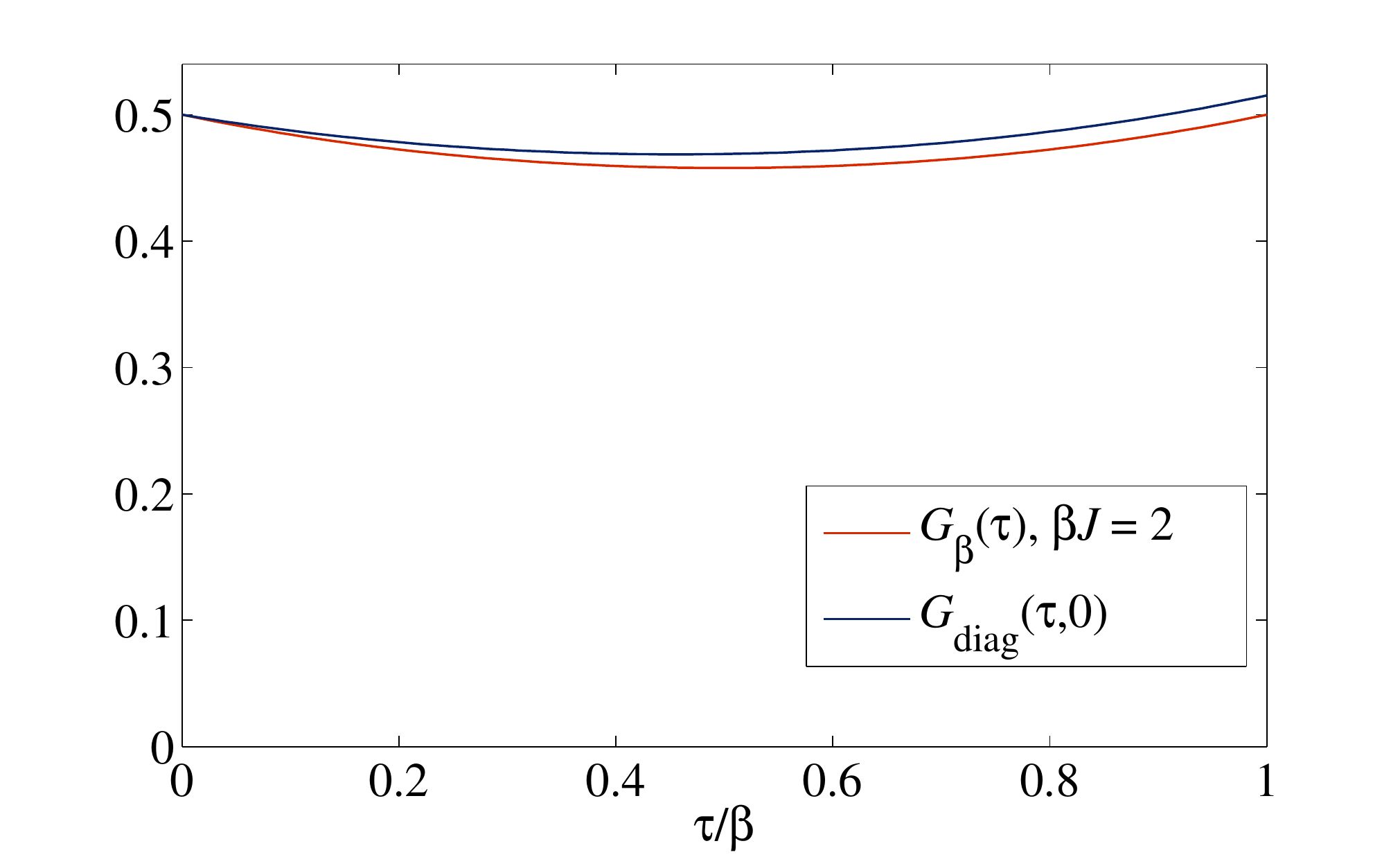}
\caption{ On the left we have $N=24$ and we plot the Euclidean time 
thermal answer and also the ratio \nref{RatioAvg}  averaged over 150 choices of the couplings. 
We also ploted the ratio \nref{RatioAvg}  for one particular value of the couplings to display how it differs from the average. 
 On the right we took $N=30$ and we show the euclidean correlator \nref{RatioNat} and the thermal one for a single realization of the couplings. 
 We see that as we increase $N$ we are approaching the result \nref{Twoptdi}. Comparing with the error for a single realization of the couplings for 
 $N=24$ (left) and $N=30$ (right)  we see that it decreases as we increase $N$.   }
\label{CouplingsAvg}
\end{center}
\end{figure}

 Here we compute the diagonal Euclidean correlators for finite $N$ and compare to the thermal answer, also 
 computed at finite $N$. 
   Here we are testing whether 
 \nref{Twoptdi} works at finite $N$. We fix an order one value of $\beta J$ so as to have contributions from a large number of states (we are not in the 
 conformal limit). For low values of $N$, such as $N=8$ we have large deviations of order 20 \%.
 However, as $N$ becomes large, we get closer results with smaller errors, see figure \nref{CouplingsAvg} for examples with $N=24, ~30$.  
  The way the error decreases seems consistent with the $1/N^3$ scaling\footnote{ In 
the left plot of \nref{CouplingsAvg} we are using the normalization in \nref{RatioAvg} where the correlator does not have to be $1/2$ at $\tau =\tau' =0$
(at finite $N$).
In the right we used the more natural normalization \nref{RatioNat} where indeed the correlator is $1/2$ at $\tau=\tau'=0$.}.

  \section{Gravity interpretation } 
  
  It is interesting to ask whether we can give a gravity interpretation to the above results. 
  At the time of writing, the full bulk dual of the SYK model is unknown. 
  However, we know that general Nearly-$AdS_2$ gravity or string theory has some features in common with 
  the low energy limit of the SYK model. In particular, in both cases we have an emergent reparametrization 
  symmetry that is both spontaneously and explicitly broken, with an explicit breaking given by the 
  Schwarzian action. 
  A simple Nearly-$AdS_2$  model is the Jackiw Teitelboim model coupled to matter, see 
  \cite{Almheiri:2014cka,Jensen:2016pah,Maldacena:2016upp,Engelsoy:2016xyb} for further discussion. 

     \begin{figure}[h!]
\begin{center}
\includegraphics[scale=.5]{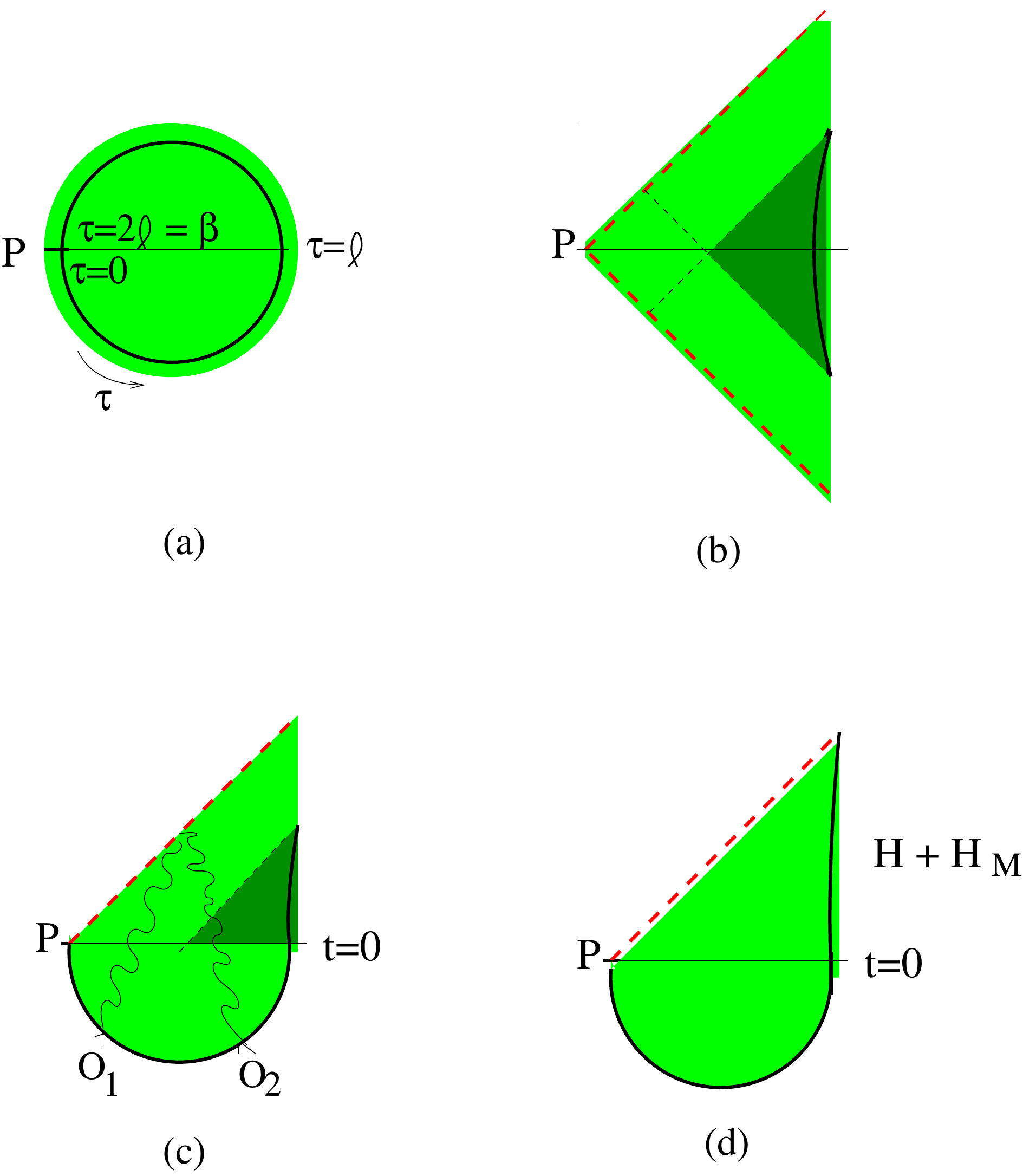}
\caption{(a) Euclidean black hole. The geometry is the  portion of $H_2$ inside the black boundary curve. 
At the point $P$ we have special boundary condition 
for the bulk fields.  (b) The corresponding Lorentzian black hole. It obtained by analytic continuation to lorentzian time along the line of reflection symmetry 
indicated by a horizontal black line passing through $P$.  Only the dark green region (color online) is visible to the boundary observer through  simple 
measurements. The boundaries of the dark green region
are the causal horizons for such observer. 
 The singular boundary conditions at $P$ create a shock wave in the bulk that follows the dotted red lines. 
 (c) By acting with operators on the boundary we can create states which are either inside or outside the horizon of figure (b). 
 (d) By changing the boundary Hamiltonian we can render visible the whole region within the doted red lines.   }
\label{BlackHoles}
\end{center}
\end{figure}

         Here we will give a qualitative gravity picture for the solutions we had above. 
   More precisely we will discuss gravity theories with similar features (symmetries) to the ones discussed above. 
   The fact that we continue to have the same diagonal correlators as the thermal state is here interpreted
    as saying that we continue
   to have the same geometry as in the Euclidean black hole. 
   In particular, at low energies, we  continue to have the same 
   $AdS_2$ geometry that we had in the thermal case. In Eulidean space we have the hyperbolic disk $H_2$ and we imagine that there
   is a boundary at some finite but very large circle.  
   See figure \ref{BlackHoles}(a). The new feature is that we have a special point labelled by $P$ in figure \ref{BlackHoles}, where we have a kind of defect. 
   We imagine that we have $N$ bulk fields and that there is a boundary condition at the marked point that relates the bulk fields in pairs. 
   For example, we can add 
     an interaction of the form $\prod_k e^{  \lambda S_k}$ for very large $\lambda$ which kills all states except the 
   ones with eigenvalue $S_k=1$ for the operators in \nref{Sk}, which are interpreted in the bulk as the product of two fermion fields 
   $i \psi^1 \psi^2$ at the corresponding bulk point. We can also imagine measuring the values of all the fields at the point $P$ on the
    boundary and projecting onto 
   the simultaneous eigenstates of all these measurements\footnote{ Complete 
   measurements on an Einstein Rosen bridge were discussed in \cite{Susskind:2014yaa,Susskind:2016jjb}. 
   }. In any case, we are doing a very high energy or UV-like insertion because it is localized at 
   one point P of the boundary circle. At all other points of the boundary circle we have the standard boundary conditions, the same as the ones we have in
   the thermal state. If we had bulk scalar fields, we can imagine that at $P$, we are putting a source for the bulk scalar fields. 
   
   This picture  has the advantage of realizing the symmetries of the problem. In fact, we can see this more clearly if we work in the so called 
   poincare coordinates,  where we send the point $P$ to infinity.
    The metric can then be written as 
   \be
   \la{Poinc} 
   ds^2_E = {dt^2 + dz^2 \over z^2 } ~,~~~~~~~~~ ds^2_L =  {-dt^2 + dz^2 \over z^2 }
   \ee
   with  Euclidean or Lorentzian signature. The Lorentzian coordinates cover the whole light (and dark) green regions of figure \ref{BlackHoles}(b). 
    On this metric we  want to consider a  field configuration or a boundary condition at large $z$ 
    that is invariant under $t$ translations. This is because this was the symmetry preserved by the 
    off diagonal low energy correlator \nref{GoffLow}. 

   We can continue this configuration to Lorentzian time in various ways. If we just continue $t \to i t $ in \nref{Poinc} we get a zero temperature configuration. 
   
   More interestingly, we can continue the metric along a moment of time reflection symmetry and obtain   the Rindler 
   $AdS_2$ coordinates,  or finite temperature solution. See figure  \ref{BlackHoles}(b). This time reflection symmetry acts as a reflection the leaves fixed the horizontal
   line passing through point $P$ in figure \ref{BlackHoles}. 
    In these coordinates the metric is the same as the one for the thermofield double, but the
   fields obey different boundary conditions which break some of the isometries of $AdS_2$. One important point is that we get a whole region behind the horizon. 
   Notice that only a portion of $AdS_2$ is visible outside the horizon, the dark green region in figure \ref{BlackHoles}(b).

   Furthermore, inserting operators in Euclidean time, we can create more general 
    states on the lorentzian $t=0$ slice. This gives us a precise way to generate states on this 
   slice. Any perturbative state can be produced by a superposition of operator insertions. 
    In particular, we can insert wavefunctions which are localized behind the horizon.  Notice that the map between operator insertions on the boundary 
    and the wavefunctions on the bulk is purely kinematical, so we can formally follow the same procedure in the SYK model to define the 
    ``bulk wavefunctions''.  In the gravity picture we can localize these excitations behind the horizon.  When we evolve in Lorentzian signature, such 
    particles will not be visible from the outside. One practical way we can try to see them, is to compute the  expectation value of the same field and 
    ask whether it can become significantly large, as it would be the case with a particle that comes out the black hole and hits the boundary. This will not
    happen for the modes localized behind the horizon if we evolve with the lorentzian SYK hamiltonian as in figure \ref{BlackHoles}(c). 
     This suggests that for each state $|B_s\rangle$ we have a picture which is similar to a gravity configuration with a smooth horizon. Of course, 
     there is a shock wave some distance behind the horizon\footnote{ The distance (or time) to the shock wave
     becomes  very small for an observer who is drops into the 
     black hole at very early times, at lorentzian times $t \ll 0$. For such observers the starting configuration is one that is complex and it is becoming simpler.
     As observed in \cite{Susskind:2015toa} such observers see some kind of firewall.  }.
     Though each of the states $|B_s\rangle$ is special, this set of special states spans the whole Hilbert space. 
            Each of these states is special because they have non-trivial correlators for the operators $S_k(t)$. 
      These non-trivial correlators decay in time as these states thermalize and become more generic, see figure \ref{Goff}. 
     

  \subsection{ Qualitative connection with other boundary state solutions } 
  
  In this subsection we consider nearly-$AdS_2$ gravity configurations containing an  end of the world brane. These are configurations 
  that can describe pure states. The end of the world ``brane'' is a particle in this case. We will see that by tuning its tension to high values 
  we get a configuration that is qualitatively similar to that shown in figure \ref{BlackHoles}(a). 
  
\begin{figure}[h]
\begin{center}
\includegraphics[scale=.5]{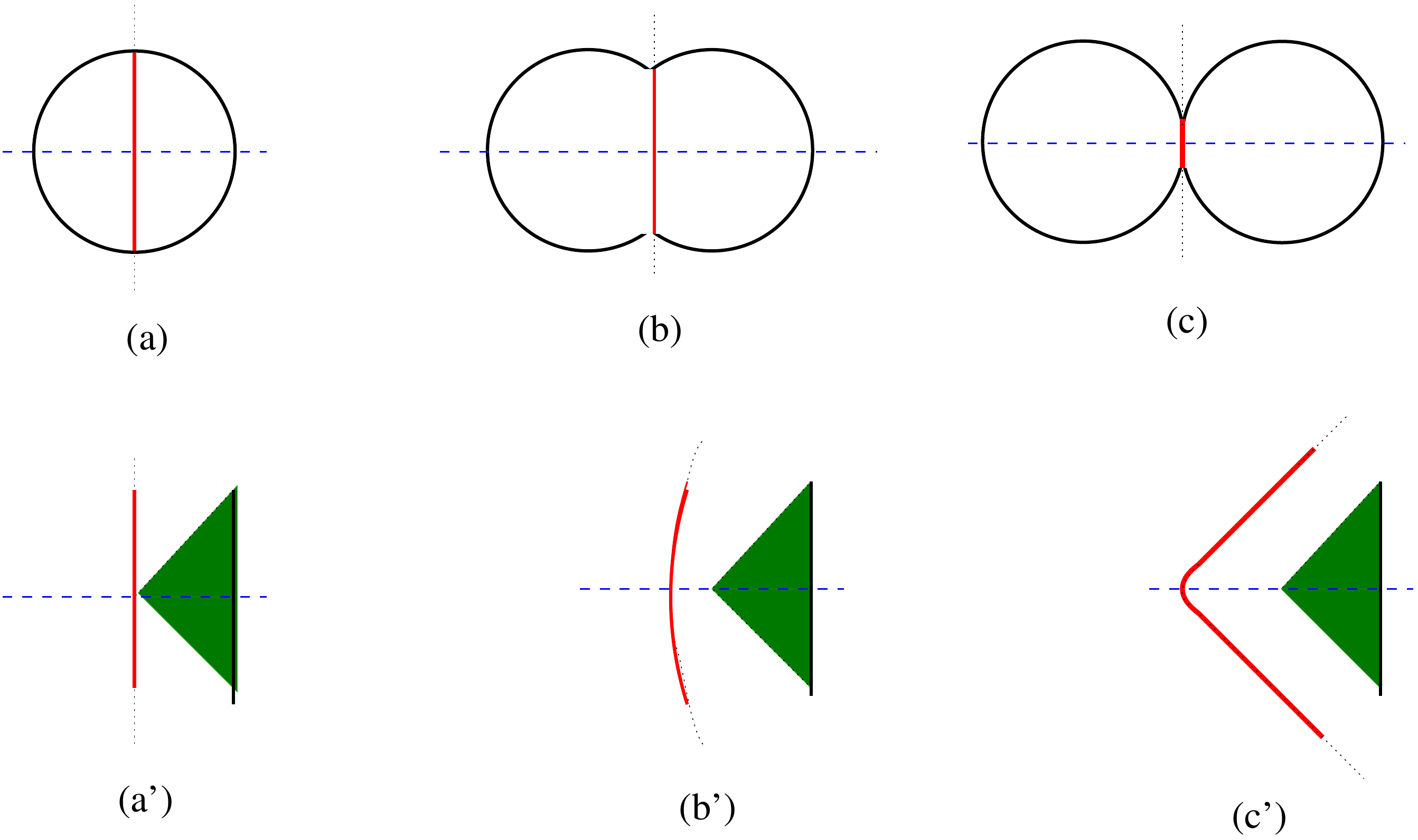}
\caption{Solutions corresponding to an end of the world particle of some mass which are obtained as a $Z_2$ quotient. (a) and (a') Euclidean and Lorentzian 
black holes in nearly-$AdS_2$ gravity. Under a $Z_2$ quotient which is generated by a reflection along the vertical red line we get a one sided black hole and 
an end of the world ``particle'' at the red line. 
(a), (b), (c) correspond to situations with larger and larger values for the   mass $\mu$.  (a') (b')  (c') are the corresponding Lorentzian solutions. 
The green triangle represents the region outside the horizon. The horizontal blue doted lines represents the moment of time reflection symmetry used
to connect the Euclidean and Lorentzian solutions. The red lines are the fixed points of the $Z_2$ reflection symmetry. The black line is the trajectory of the
boundary in $AdS_2$ space.  
 }
\label{CircleBrane}
\end{center}
\end{figure}

  Let us first remind the reader that a simple way to generate a gravity solution dual to a pure state is to take the eternal $AdS$ Schwarzschild solution 
  and perform a $Z_2$ quotient that exchanges the left and the right sides, a reflection along a vertical line in the Penrose diagrams, see figure 
  \ref{CircleBrane}(a,a').  Whether we can or cannot do this quotient depends on the full gravity theory in question. In some examples that arise from 
  string theory we can certainly do it and the end of the world branes are the ones familiar in string constructions, see e.g. \cite{Maldacena:2001kr}.
  We will not discuss the full UV complete gravity theory here, we will simply consider a phenomenological model where we have nearly-$AdS_2$ gravity and
  we have an end of the world particle with an arbitrary mass $\mu$. We generate these configurations by going to a covering space containing a 
  particle of mass $\mu$ and then doing a $Z_2$ quotient, where the particle sits as the $Z_2$ invariant points. 
  We will see that, 
     as we increase the mass,
 the effects of gravitational backreaction move this end of the world brane deeper into the left side of the black hole geometry until we get a picture rather similar
 to the one in figure \ref{BlackHoles}. We will now discuss this in more detail.

 As explained in \cite{Maldacena:2017axo}, the gravitational dynamics in nearly-$AdS_2$ gravity can be expressed in terms of the motion of a boundary 
 particle in a rigid $AdS_2$\footnote{ Do not confuse the boundary of $AdS_2$ with the actual physical boundary which sits at the location of the boundary particle. 
 The boundary particle is in the interior of $AdS_2$ but far away from the central region of interest \cite{Maldacena:2017axo}.}. Therefore, we are looking for a $Z_2$ invariant configuration that contains the boundary particle and the  particle of mass $\mu$
  going between 
 two points on the boundary particle trajectory.
  The  boundary particle emits the massive bulk particle and it absorbes it again later, see figure \ref{CircleBrane}. 
 Computing the classical solution including backreaction amounts to finding particle trajectories of this kind so that the energy momentum is conserved at the 
 vertices. Each of the particle trajectories is specified in terms of their $SL(2)$ charges, and the energy momentum conservation amounts to the requirement that
 the total $SL(2)$ charge is zero \cite{Maldacena:2017axo}. 
 It is convenient to describe $AdS_2$ in terms of embedding coordinates $Y^a = (Y^{-1},Y^0,Y^1)$, $Y.Y = - ( Y^{-1} )^2 - (Y^0)^2 + (Y^1)^2 =-1$. 
 We can also view the $SL(2)$ charges associated to the particle trajectories as a vector $Q^a$. We can pick the charge for the massive particle as 
 \be
  Q^a_\mu = ( 0,0,\mu)~,~~~~~~~~ Y.Q =0
  \ee
  where we have also written the equation for the geodesic trajectory. 
  The boundary particles do not follow geodesics, they move as if they where charged particles of charge $q$ and mass $m$ 
   in a uniform electric field in $AdS_2$\footnote{ In terms of the parameters of the nearly-$AdS_2$ gravity theory we have $q = 2 \Phi_b = 2\Phi_r/\epsilon$
   and $m = q - \epsilon M$, where $M$ is the ADM mass of the black hole,  and we imagine taking $\epsilon \to 0 $ \cite{Maldacena:2017axo}.}.
  Their trajectories are instead given by \cite{Maldacena:2017axo}
  \be \la{Traj}
   Y .Q_R = - q  ~,~~~~~~~Y.Q_L = +q   ~~,~~~~~~~Q^2 = m^2 -q^2
   \ee
   The trajectories specified by \nref{Traj} look like circles in Euclidean space where the center is at $Y^a \propto Q^a$.
   The center of these circles is 
   also the position where the future and past horizons intersect in the corresponding Lorentzian black hole.  
  For the problem 
   we are interested in, we expect that these circles will be displaced as in figure \ref{CircleBrane}(b,c). 
   Therefore we make the ansatz 
   \be
   Q^a_R = A ( \cosh \rho_0 , 0, \sinh \rho_0) ~,~~~~~~~ Q^a_L = - A ( \cosh \rho_0 , 0, - \sinh \rho_0 )  ~,~~~~A^2 = q^2 -m^2
   \ee
   where we can view $\rho_0$ as the displacement of the center of the circle relative to the origin, $Y^0=Y^1=0$. 
   Demanding that the sum of the charges is zero $Q^a_R + Q^a_L + Q_\mu^a=0$ we find that 
   \be \la{muva}
   \mu = 2 A \sinh \rho_0
   \ee
   But due to \nref{Traj} we find that the size of each circular trajector is given by $A \cosh \rho_c = q$. In order for $\rho_c$ to be larger than 
   $\rho_0$ in 
   \nref{muva} we need that 
   \be
    \mu < 2 m 
    \ee
    Which is clear if we think in terms of the balance of forces at the vertex of figure \ref{CircleBrane}(c). The point is that as $\mu \to 2 m$ the two circles become 
    almost tangent to each other, and after the $Z_2$ quotient we get a geometry qualitatively similar to that in figure \ref{BlackHoles}(a). 
   This is a large value for $\mu$, which is
    comparable to the size of the UV cutoff where the boundary particle sits. 
    
    The analytic continuation of these configurations is such that we get a geometry that contains a horizon and an end of the world particle inside the horizon. 
    The trajectory of this particle is following a geodesic in the ambient $AdS_2$ spacetime. As we move from figure \ref{CircleBrane}(a') to \ref{CircleBrane}(c') this
    is a geodesic that starts closer and closer to the boundary. So as we get to figure \ref{CircleBrane}(c') it looks like a shock wave.

  \section{ Evolution of the state under a different Hamiltonian}

  Previously we have claimed that if we start 
  with the state $|B(\ell) \rangle$   at $t=0$ and evolve it in Lorentzian time, then 
  we get a state that is similar to the gravity configurations in \nref{BlackHoles}(b,c) which contain a horizon and an
   inaccessible region behind it. 
  
  Here we will show that by evolving with a different Hamiltonian, one that is fine tuned to the particular state $|B_s\rangle$ that used to prepare the 
  initial state, then we can get a modified evolution for the Schwarzian degree of freedom. Interpreted  as a statement in 
  nearly-$AdS_2$ gravity, this modified evolution is such that it allows us to see behind the horizon. A related modified evolution involving 
  a double trace deformation for pure black holes in $AdS_3$ is being considered by \cite{Ahmed}. 
    
   This is done as follows. We add to the SYK Hamiltonian \nref{Ham} a new term $H_{M}$ of the form 
   \be \la{TotHam}
    H_{\rm total} = H_{SYK} + \epsilon   H_M ~,~~~~~~H_M = - J \sum_{k=1}^{N/2} s_k  i \psi^{ 2 k-1} \psi^{ 2k } 
    \ee
    here $H_M$ is an operator which looks like a ``mass term'' for the fermions.  The factor of $J$ in $H_M$ is setting the units so that that $\epsilon $ is dimensionless\footnote{
The SYK model plus a mass term was discussed in \cite{Ferrari,Garcia-Garcia:2017bkg}.}. 
     It is very important for our discussion that we choose the signs $s_k$ in \nref{TotHam} to be the same as the ones that describe the state $|B_s (\ell) \rangle$
     \nref{Bs}.

           At large $N$ we could analyze this problem by solving the real time Schwinger-Dyson equations for the full Hamiltonian \nref{TotHam}. 
           This was done in \cite{Eberlein:2017wah} for a closely related situation. 
          Here we will further  assume that $\epsilon$ is small enough so that we can first solve the SYK problem
           and then treat $\epsilon$ as a small perturbation that will affect
    only the low energy dynamics of the model. At low energies this dynamics is dominated by the reparametrization mode $f$ and we will consider only the
    effects of this mode. 
     
    In other words we simply evaluate the extra term, $H_M$, on the original state and integrate over reparametrizations 
    \bea \la{MassExp}
   &~& \langle e^{ - i \int dt H_M(t) } \rangle \sim \int {\cal D} f  e^{ i S[f] - i \int dt \langle H_{M}(f(t) ) \rangle  }  ~,~~~~ 
   \cr
   &~&  \langle B_s (\ell) | H_M(t) |  B_s (\ell) \rangle \sim - N J G_{\rm off}(t,t) \sim - 
   2 N { J c_\Delta^2  \over \left[{ J \beta  \over \pi }   \cosh{ \pi t \over \beta } \right]^{4 \Delta } }  
    \eea
    We now couple the reparametrization mode by  transforming $G_{\rm off}$ in this expression by a reparametrization as in \nref{ReparOff}. 
   \bea
    S_M[\varphi] & =&  2 \epsilon J c_\Delta^2 N 
    \int d t { [ \varphi'(t)]^{ 2 \Delta }  \over \left[ { \beta J  \over \pi } \cosh{ \pi \varphi(t) \over \beta } \right]^{4 \Delta } } = 2 \epsilon J c_\Delta^2 N   \int dt   (f')^{2\Delta } 
    \la{MassTe}
    \eea
    where 
    \be
    f = { \pi \over J^2 \beta }  \tanh { \pi \varphi(\tau) \over \beta }   
    \ee
     
    By introducing a Lagrange multiplier, $\lambda(t)$,  it is possible to write the total action, which is the Schwarzian \nref{Sch}  plus  \nref{MassTe}, as 
    \be
    S_{\rm tot} = N  { \alpha \over 2 } \int dt  \left[  { 1 \over J } { \dot \phi^2   } + \lambda ( e^\phi - f' ) +  \hat \epsilon J  e^{ 2 \Delta \phi  }  \right] ~,~~~~~
    \hat \epsilon \equiv     { 4 c_\Delta^2 
  \epsilon \over \alpha }
    \ee
    We set  $ \hat  \epsilon$ to zero while we do the Euclidean evolution to prepare the state $|B_s(\ell) \rangle$ 
    and we can turn it on   as we start the Lorentzian time evolution at $t=0$. 
    See figure \ref{BlackHoles}(d). 
    The Euclidean time solution we are interested in is $f = { \pi \over J^2  \beta } \tan { \pi \tau \over \beta } $, 
    which sets $\lambda =  -    J$.$ ~~~~ $\footnote{
    The Euclidean time $\tau$ here  is related to the euclidean time in previous sections via $\tau_{\rm previous} = { \beta \over 2} + \tau_{\rm here}$.}. 
    Since the equation of motion for $f$ implies that $\lambda $ is constant, we can keep the same constant after we start the Lorentzian evolution. 
    
\begin{figure}[h]
\begin{center}
\includegraphics[scale=.45]{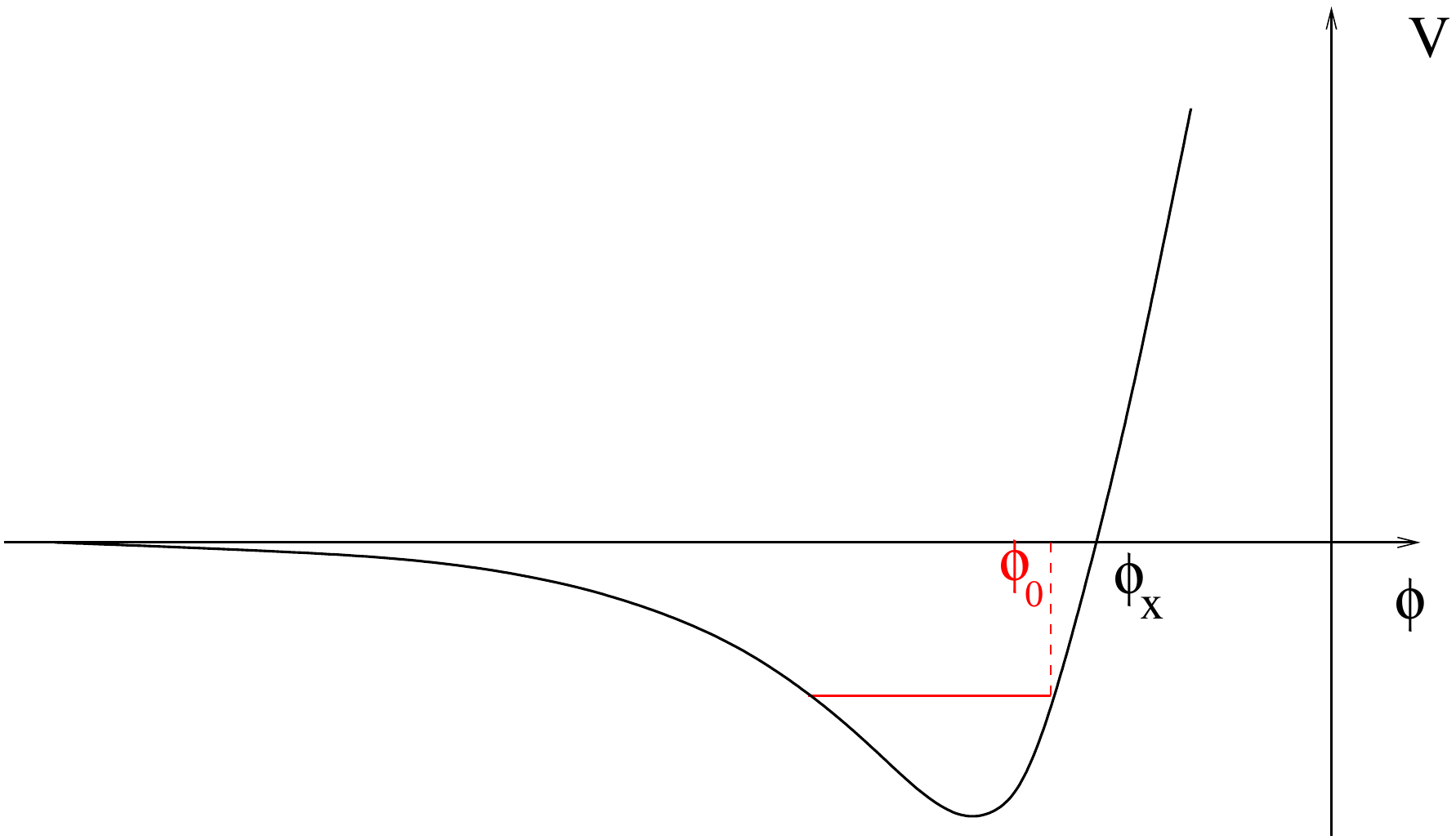}
\caption{ Sketch of the potential \nref{Pot} that determines the modified evolution. If the initial position, $\phi_0$, is less than $\phi_\times$, then $\phi$ oscillates
in the allowed region, given by the red line here. Since $\phi \sim \log z $ we see that $z$ in \nref{Poinc} does not approach zero at late times. 
 }
\label{Potential}
\end{center}
\end{figure}

 The Lorentzian evolution is simply the motion of a particle with coordinate $\phi$ on a potential 
    \be \la{Pot}
    V = - \lambda e^{ \phi }  - \hat  \epsilon J e^{ 2 \Delta  \phi } = J [    e^\phi - \hat  \epsilon e^{ 2 \Delta \phi } ]   
    \ee 
    This potential crosses zero at $\phi = \phi_\times$, given by 
    \be
      e^{ (1- 2 \Delta ) \phi_\times } = { \hat\epsilon    } ~,~~~~{\rm  for }~~~ ~~ 0 < \Delta < 1/2
    \ee
    The potential is negative for $\phi < \phi_\times$ and positive for $\phi> \phi_\times $. See figure \ref{Potential}.

          Furthermore at $t=0$, $\dot \phi =0$, since it is a moment of time reflection symmetry. And the value of $\phi$ at $t=0$ is given by 
          \be
          e^{\phi_0} = ( f')_{t=0} = \left( {\pi \over J^2 \beta } \tanh{ \pi t \over \beta} \right)'_{t=0} =  { \pi^2 \over (\beta J)^2  }
          \ee
          We see that as long as $\phi_0 < \phi_\times $, then the subsequent motion for $\phi$ remains bounded. 
         We will obey this condition as long as $\epsilon$ is large enough 
         \be \la{CondEp}
         \phi_0 < \phi_\times  \longrightarrow   \left(  { \pi \over \beta J  } \right)^{2  (1-2 \Delta) } < { \hat \epsilon  }  \ll 1 
         \ee
         The last inequality comes from the condition that we can trust 
        the reduction of the dynamics  to the reparametrization mode. We can obey both conditions in \nref{CondEp}  as long as $\beta J \gg 1$.

    In terms of the $AdS_2$ coordinates \nref{Poinc} we know that $t \propto f$ and $ z \propto f' = e^{ \phi }  $.
     Therefore we see that the motion of the boundary along the $z$ direction is  oscillatory but  bounded. In particular, 
     it  does not approach $z=0$ (or $\phi \to - \infty $).  This means
    that the region that is visible from the boundary is the whole Poincare patch. See figure \ref{BlackHoles}(d).  
        
  \subsection{ Adding particles behind the horizon} 
   
   As we discussed above, we can add operators during the Euclidean evolution in order to add particles to the original background.  
   We know that we can represent any wavefunction on top of the Hartle-Hawking vacuum by adding a suitable superposition of operators inserted in the 
   lower part of the Euclidean background. Therefore, we can add particles that are either outside or inside the horizon, see figure \ref{BlackHoles}(c). 
   If we evolve with the original SYK Hamiltonian, the same Hamiltonian we used to prepare the state during Euclidean evolution, then the particles behind the 
   horizon are not visible. On the other hand, if we evolve the state with the modified Hamiltonian, then all particles become visible.

  \subsection{Relation to traversable wormholes} 
 
  The addition to the Hamiltonian \nref{TotHam} is similar to the one that was considered in the problem of
  traversable wormholes \cite{Gao:2016bin,Maldacena:2017axo}\footnote{In \cite{Maldacena:2017axo} the interaction was turned on only at one time. Here it is
  turned on for all $t\geq 0$.}. 
  The boundary state $|B_s\rangle$ is obtained by measuring all the spins on the left  half of the thermofield double, giving a result $\{ s_k\}$. With this 
  knowledge, an observer on the right half can  act   with an operator that exploits these results, as in \nref{TotHam},
    and access a larger region of the spacetime. With the interaction 
  considered above, the observer can access the whole region in the Poincare patch instead of just the Rindler patch. 
    Note that the measurement is responsible for creating the shock wave in the bulk. 
  
  \section{ Discussion } 
  
  In this paper we have looked at a particular set of simple microstates of the SYK model. 
  They are defined by a simple boundary condition for the Majorana fermions. 
  They are also joint eigenvectors of a set of commuting operators $S_k$. 
  These simple states span a complete basis of the Hilbert space of the model. 
   We have further projected them into a lower energy subspace by performing some amount $\ell$ of Euclidean time evolution. 
   
   We found that the ``diagonal'' correlators are exactly given by the thermal ones. 
   We interpret this as saying that these simple states look completely thermalized from the point of view of the diagonal 
   correlators. This also implies that the expectation values of operators that appear in the $\psi^i \psi^i$ OPE at order $1/N$ also have the same
   expectation values as in the thermal state. This is similar, in spirit, to the fact that these correlatorss for a single value
   of the couplings is the same, at large $N$, as the average over couplings. Here the correlators for a single state
   looks similar to the thermal state, which is the average over all states. 
   Some particular off diagonal correlators are not thermalized and   know the details about the particular state $|B_s\rangle$. As we evolve in Lorentzian time
   this information is effectively lost as the state thermalizes. 
   
   We have discussed some nearly-$AdS_2$ phenomenological gravity theories that have similar properties. 
   In these gravity theories, the state is a configuration with only one boundary which  contains some kind of 
   end of the world particle in the interior. The geometry contains a spacetime region    that is causally inaccessible
   from the outside. In the Lorentzian solution this end of the world particle starts in a region close to where
   the left boundary would be in in the full wormhole solution and it quickly becomes a high energy shock wave which is
   at some distance behind the horizon.  Observers who fall into the black hole from the right side at positive times
   experience a smooth horizon. 
   
   In the SYK model it is possible to change the Hamiltonian evolution so that the main effects can be captured 
   by the Schwarzian action plus a contribution  induced by the extra term in the Hamiltonian. This modification of the 
   evolution for  the reparametrization goldstone boson  can be interpreted in the bulk as a new trajectory for the boundary. 
   This new trajectory is such that it is possible to see the whole spatial slice of the original state. In particular, we 
   can see the region that previously was behind the horizon. This effect is basically the same as the one 
   that makes wormholes traversable. In fact, there is a precise connection between the two. In the SYK model, 
   we can view the initial state as the one that results from measuring a complete set of commuting operators, 
   $S_k$ on the left side of a thermofield double. Then the modified evolution is essentially the same as the one
   that was considered in the context of traversable wormholes \cite{Gao:2016bin,Maldacena:2017axo}. 
   
   It is tempting to conjecture that for more general cases, such as the black hole dual to the D0 brane matrix model \cite{Itzhaki:1998dd} 
   (or BFSS matrix model \cite{Banks:1996vh}) 
   we have a similar picture. Namely, that a full measurement in the 
   ``simple'' basis 
   of the UV state at $\tau=0$ on the left side of the thermofield double state is represented by some operator which is inserted at point $P$ in 
   figure \nref{BlackHoles}, so as to give a smooth horizon configuration. It would be nice to check this.

{\bf Acknowledgements } 

We thank A. Almheiri, D. Stanford and Z. Yang for discussions. The thank S. Sachev for a pre-publication copy of \cite{Eberlein:2017wah}. 
J.M. is supported in part by U.S. Department of Energy grant
de-sc0009988. I.K. thanks the Institute for Advanced Studies for hospitality
while this work was being finished.

\mciteSetMidEndSepPunct{}{\ifmciteBstWouldAddEndPunct.\else\fi}{\relax}
\bibliographystyle{utphys}
\bibliography{QuenchesDraft.bib}{}

\end{document}